\documentstyle[prl,aps,epsfig,multicol]{revtex}
\newcommand{\beq}{\begin{equation}}
\newcommand{\eeq}{\end{equation}}

\begin{document}

\title{$\lambda /4$, $\lambda /8$, and higher order atom gratings via Raman
transitions}
\author{B. Dubetsky and P. R. Berman}

\address{Michigan Center for Theoretical Physics, FOCUS Center, and Physics\\
Department, University of Michigan, Ann Arbor, MI 48109-1120}

\date{\today}
\maketitle

\begin{abstract}
A method is proposed for producing atom gratings having period $\lambda /4$
and $\lambda /8$ using optical fields having wavelength $\lambda $.
Counterpropagating optical fields drive Raman transitions between ground
state sublevels. The Raman fields can be described by an effective two
photon field having wave vector $2{\bf k}$, where ${\bf k}$ is the
propagation vector of one of the fields. By combining this Raman field with 
{\em another} Raman field having propagation vector $-2{\bf k}$, one, in
effect, creates a standing wave Raman field \label{91}
whose ``intensity'' varies as $\cos (4{\bf k\cdot r}).$ When atoms move
through this standing wave field, atom gratings having period $\lambda /4$
are produced, with the added possibility that the total ground state
population in a given ground state manifold can have $\lambda /8$
periodicity. The conditions required to produce such gratings are derived.
Moreover, it is shown that even higher order gratings having periodicity
smaller than $\lambda /8$ can be produced using a multicolor field geometry
involving three (two-photon) Raman fields. Although most calculations are
carried out in the Raman-Nath approximation, the use of Raman fields to
create reduced period optical lattices is also discussed.
\end{abstract}

\section{Introduction}

The field of optical coherent transients \cite{berrev} had at its origin the
pioneering work of Kurnit, Abella, and Hartmann on photon echoes \cite
{phoecho}. Since that work, Hartmann and coworkers have made a large number
of significant contributions in both cw and transient nonlinear
spectroscopy. One of us (PRB) had the good fortune to collaborate with his
group on problems related to diffractive scattering of atoms in
superposition states \cite{diff}. There were also a number of spirited
discussions on the billiard ball model of photon echoes \cite{billiard}, a
model which was somewhat before its time since it is now used commonly in
theoretical models of atom interferometers \cite{ai}. It is with great
pleasure that we participate in this {\em festschrift} volume to honor Sven
Hartmann.

The first wave of experiments on optical coherent transients, such as photon
echoes and free precession \cite{fp}, involved coherence between different
electronic states that are coupled by an optical field. Subsequent
experiments, however, such as the {\em stimulated photon echo }\cite{spe}%
{\em , }employed optical fields as a means for creating coherence between
ground state sublevels (or a spatial modulation of a single ground state
population). Since ground state coherences can have decoherence times
approaching seconds, rather than the tens or hundreds of nanoseconds
associated with optical transitions, ground state coherence has important
applications in atom interferometry and quantum information. There is now an
extensive literature on both {\em cw} and coherent transient, ground state
spectroscopy.

In recent years, both {\em cw }and pulsed optical fields have been used to
control the center-of-mass motion of atoms. Of particular interest to the
current discussion is the possibility of using optical fields having
wavelength $\lambda $ to create {\em high-order matter wave gratings} having
period $\lambda /2n$, where $n$ is a positive integer greater than or equal
to $2$ \cite{highorder}. Such matter wave gratings could serve as efficient
scatters of soft x-rays. There have been several schemes proposed for
achieving this goal, but most of these schemes involve several atom-field
interaction zones. Recently we showed that it is possible to create
high-order gratings in a {\em single} interaction zone using a set of
counterpropagating optical fields having different frequencies \cite{ho}.
One of the limitations of that work was set by the lifetime of the excited
state of the transition. The product of atom-field detunings and excited
state lifetime has to be much greater than unity to satisfy certain
adiabaticity requirements, and this necessitates the use of large detunings
that leads to a corresponding decrease in transition amplitudes.

These problems can be avoided if one replaces the optical transitions with
ground-state, Raman transitions. A pair of fields drives transitions from
one ground state sublevel to another and acts as an effective two-photon
field that is the analogue of the single field that drives an optical
transition. In analogue with the optical case, the two-photon field leads to
a spatially modulated Raman coherence, but {\em not} to a spatially
modulated population of the ground state sublevels\cite{singlecoupling}.
However, a {\em pair }of such two-photon fields, acting as a {\em Raman
standing wave field}, leads to ground state gratings having period $\lambda
/4$. With a proper choice of field polarization and initial conditions, the
population density in these systems can have period $\lambda /8$, even if
the overall period of the matter wave gratings (including the dependence on
internal states) has periodicity $\lambda /4$. Moreover, a field geometry
involving three pairs of fields can be used to produce gratings having
period $\lambda /4n$, where $n$ is a positive integer greater than 2. We
first describe the method for producing $\lambda /4$ periodicity and then
discuss briefly how to extend this method to produce higher-order gratings.
The possibility of creating optical lattices having period $\lambda /4n$
using these techniques is explored.

\section{Basic Formalism}

Consider an atom interacting with a field that consists of a sum of
traveling wave fields. The total electric field vector is given by 
\begin{equation}
{\bf E}\left( {\bf r,}t\right) =\frac{1}{2}E_{j}{\bf e}_{j}\exp \left( i{\bf %
k}_{j}\cdot {\bf r}-i\Omega _{j}t\right) +c.c.,  \label{1}
\end{equation}
where $E_{j},$ ${\bf e}_{j},$ ${\bf k}_{j}\ $and $\Omega _{j}$ are the
amplitude, polarization, wave vector and frequency of wave $j$; there is a
summation convention implicit in Eq. (\ref{1}) and below in which repeated
indices appearing on the right-hand side of an equation are to be summed
over, except if these indices also appear on the left-hand side of the
equation. The field (\ref{1}) drives transitions between a ground-state
manifold characterized by quantum numbers $L_{G}$ (total orbital angular
momentum), $S$ (total spin angular momentum), $J_{G}$ (coupling of $L_{G}$
and $S_{G}$), $I$ (total nuclear spin angular momentum), $G$ (coupling of $%
J_{G}$ and $I$), $m_{g}$ (Zeeman quantum number), and excited-state manifold
characterized by quantum numbers $L_{H}$, $S,$ $J_{H},$ $I,$ $H,$ $m_{h}.$
In the resonant or rotating-wave approximation the atomic state probability
amplitudes evolve as 
\begin{mathletters}
\label{2}
\begin{eqnarray}
i\dot{a}_{Hm_{h}} &=&\exp \left( -i\Delta _{Hm_{h},Gm_{g}}^{\left( j\right)
}t+i{\bf k}_{j}\cdot {\bf r}\right) \chi _{Hm_{h},Gm_{g}}^{\left( j\right)
}a_{Gm_{g}}-\gamma _{H}a_{Hm_{h}}/2,  \label{2a} \\
i\dot{a}_{Gm_{g}} &=&\exp \left( i\Delta _{Hm_{h},Gm_{g}}^{\left( j\right)
}t-i{\bf k}_{j}\cdot {\bf r}\right) \left[ \chi _{Hm_{h},Gm_{g}}^{\left(
j\right) }\right] ^{\ast }a_{Hm_{h}},  \label{2b}
\end{eqnarray}
where 
\end{mathletters}
\begin{equation}
\chi _{Hm_{h},Gm_{g}}^{\left( j\right) }=\left( -1\right) ^{\nu }\chi
_{HG}^{\left( j\right) }e_{-\nu }^{\left( j\right) }\left( 2H+1\right)
^{-1/2}\left\langle Gm_{g},1\nu |Hm_{h}\right\rangle ,  \label{3}
\end{equation}
is a Rabi frequency, 
\[
\Delta _{Hm_{h},Gm_{g}}^{\left( j\right) }=\Omega _{j}-\omega
_{Hm_{h},Gm_{g}} 
\]
is an atom-field detuning, 
\[
\chi _{HG}^{\left( j\right) }=-\mu _{HG}E_{j}/2\hbar 
\]
is a reduced Rabi frequency$,$ $\mu _{HG}$ is the reduced matrix element of
the dipole moment operator between states $H$ and $G$, $\gamma _{H}$ is an
excited state decay rate, $e_{\nu }^{\left( j\right) }$ are spherical
components of the polarization ${\bf e}_{j}$, 
\begin{equation}
e_{\pm 1}^{\left( j\right) }=\mp \frac{e_{x}^{\left( j\right) }\pm
e_{y}^{\left( j\right) }}{\sqrt{2}}\text{; \ \ }e_{0}^{\left( j\right)
}=e_{z}^{\left( j\right) }  \label{31}
\end{equation}
and{\bf \ }$\left\langle Gm_{g},1\nu |Hm_{h}\right\rangle $ is a
Clebsch-Gordan coefficient.

For far detuned fields $\left( \left| \Delta _{Hm_{h},Gm_{g}}^{\left(
j\right) }\right| \gg \gamma _{H},\left| \chi _{HG}^{\left( j\right)
}\right| \right) $, one can use a secular approximation, where excited
states amplitudes adiabatically follow ground state amplitudes as 
\begin{equation}
a_{Hm_{h}}=\exp \left( -i\Delta _{Hm_{h},Gm_{g}}^{\left( j\right) }t+i{\bf k}%
_{j}\cdot {\bf r}\right) \left( \chi _{Hm_{h},Gm_{g}}^{\left( j\right)
}/\Delta _{Hm_{h},Gm_{g}}^{\left( j\right) }\right) a_{Gm_{g}}.  \label{4}
\end{equation}
Substituting this expression in Eq. (\ref{2b}) one arrives at the
Schr\"{o}dinger equation for the ground state manifold 
\begin{equation}
i\dot{a}_{G^{\prime }m_{g}^{\prime }}=\left\langle G^{\prime }m_{g}^{\prime
}\left| V\right| Gm_{g}\right\rangle a_{Gm_{g}},  \label{5}
\end{equation}
where $\hbar V$ is a reduced Hamiltonian with matrix elements 
\begin{equation}
\left\langle G^{\prime }m_{g}^{\prime }\left| V\right| Gm_{g}\right\rangle
=\exp \left[ -i\delta _{G^{\prime }m_{g}^{\prime },Gm_{g}}^{\left(
jj^{\prime }\right) }t+i{\bf k}_{jj^{\prime }}\cdot {\bf r}\right]
A_{G^{\prime }G}^{\left( jj^{\prime }\right) }\left( K\right) \left\langle
G^{\prime }m_{g}^{\prime },KQ|Gm_{g}\right\rangle \varepsilon _{Q}^{K}\left(
jj^{\prime }\right) ,  \label{6}
\end{equation}
\begin{equation}
\delta _{G^{\prime }m_{g}^{\prime },Gm_{g}}^{\left( jj^{\prime }\right)
}=\Omega _{j}-\Omega _{j^{\prime }}-\omega _{G^{\prime }m_{g}^{\prime
},Gm_{g}}  \label{7}
\end{equation}
is the Raman detuning associated with the two-quantum transition $%
Gm_{g}\rightarrow G^{\prime }m_{g}^{\prime }$ involving absorption from
field $j$ and emission into field $j^{\prime }$, 
\begin{equation}
A_{G^{\prime }G}^{\left( jj^{\prime }\right) }\left( K\right) =\left(
-1\right) ^{G^{\prime }+H+K}\left[ \chi _{HG}^{\left( j\right) }\left( \chi
_{HG^{\prime }}^{\left( j^{\prime }\right) }\right) ^{\ast }/\Delta
_{H,G}^{\left( j\right) }\right] \left[ \left( 2K+1\right) /\left(
2G+1\right) \right] ^{1/2}\left\{ 
\begin{array}{ccc}
G & G^{\prime } & K \\ 
1 & 1 & H
\end{array}
\right\} ,  \label{8}
\end{equation}
$\left\{ \ldots \right\} $ is a 6-J symbol, and 
\begin{equation}
\varepsilon _{Q}^{K}\left( jj^{\prime }\right) =\left( -1\right) ^{\nu
^{\prime }}e_{\nu }^{\left( j\right) }\left( e_{-\nu ^{\prime }}^{\left(
j^{\prime }\right) }\right) ^{\ast }\left\langle 1\nu ,1\nu ^{\prime
}|KQ\right\rangle  \label{9}
\end{equation}
is a tensor coupling vectors ${\bf e}_{j}$ and ${\bf e}_{j^{\prime }}.$ It
has been assumed that single photon detuning is much larger than any Zeeman
splittings, i.e. $\Delta _{Hm_{h},Gm_{g}}^{\left( j\right) }\approx \Delta
_{H,G}^{\left( j\right) }.$

If the single photon detunings are much larger than the excited state
hyperfine splitting, $\Delta _{H,G}^{\left( j\right) }\approx \Delta
_{J_{H},G}^{\left( j\right) }$, one can sum over $H$ to arrive at 
\begin{eqnarray}
A_{G^{\prime }G}^{\left( jj^{\prime }\right) }\left( K\right) &=&\left(
-1\right) ^{G^{\prime }+J_{G^{\prime }}+J_{G}+J_{H}+I}\left[ \chi
_{J_{H}J_{G}}^{\left( j\right) }\left( \chi _{J_{H}J_{G^{\prime }}}^{\left(
j^{\prime }\right) }\right) ^{\ast }/\Delta _{J_{H},G}^{\left( j\right) }%
\right] \left[ \left( 2K+1\right) \left( 2G^{\prime }+1\right) \right] ^{1/2}
\nonumber \\
&&\times \left\{ 
\begin{array}{ccc}
J_{G} & J_{G^{\prime }} & K \\ 
1 & 1 & J_{H}
\end{array}
\right\} \left\{ 
\begin{array}{ccc}
J_{G^{\prime }} & I & G^{\prime } \\ 
G & K & J_{G}
\end{array}
\right\} .  \label{10}
\end{eqnarray}
Finally, if\ the single photon detunings are larger than the excited state
fine structure intervals, $\Delta _{J_{H},G}^{\left( j\right) }\approx
\Delta _{L_{H},G}^{\left( j\right) },$ one finds 
\begin{eqnarray}
A_{G^{\prime }G}^{\left( jj^{\prime }\right) }\left( K\right)  &=&\left(
-1\right) ^{G^{\prime }+J_{G^{\prime }}+J_{G}+I+S+L_{H}+K}\left[ \chi
_{L_{H}L_{G}}^{\left( j\right) }\left( \chi _{L_{H}L_{G}}^{\left( j^{\prime
}\right) }\right) ^{\ast }/\Delta _{L_{H},G}^{\left( j\right) }\right] \left[
\left( 2K+1\right) \left( 2G^{\prime }+1\right) \left( 2J_{G}+1\right)
\left( 2J_{G^{\prime }}+1\right) \right] ^{1/2}  \nonumber \\
&&\times \left\{ 
\begin{array}{ccc}
L_{G} & S & J_{G^{\prime }} \\ 
J_{G} & K & L_{G}
\end{array}
\right\} \left\{ 
\begin{array}{ccc}
L_{G} & L_{G} & K \\ 
1 & 1 & L_{H}
\end{array}
\right\} \left\{ 
\begin{array}{ccc}
J_{G^{\prime }} & I & G^{\prime } \\ 
G & K & J_{G}
\end{array}
\right\} ,  \label{11}
\end{eqnarray}
In the latter case, for alkali metal atoms having $L_{G}=0,$ only $K=0$
contributes to the sum. As a consequence, one cannot couple different ground
state sublevels if the single photon detunings are larger than excited state
fine structure splitting. The two-photon Raman field acts as a scalar in
this case.

\section{Atom gratings}

Equations (\ref{5}), with the effective Hamiltonian (\ref{6}) are the
starting point for a wide class of problems involving both $cw$ and
transient ground state spectroscopy. If one diagonalizes the effective
Hamiltonian (\ref{6}), he obtains the spatially-dependent optical potentials
that characterize this atom field interaction. We will return to this point
in the Discussion. It has been assumed implicitly in Eq. (\ref{5}) that $%
{\bf r}$, the atomic center-of-mass coordinate, is a classical variable;
however, it is possible to generalize these equations to allow for
quantization of the center-of-mass motion by addition of a term $\hbar
\nabla ^{2}a_{G^{\prime }m_{g}^{\prime }}/2M$ ($M$ is an atomic mass) to the
left hand side of Eq. (\ref{5}). In this paper, we analyze problems in which
atoms are subjected to a radiation pulse whose duration is sufficiently
short to justify neglect of this kinetic energy term (Raman-Nath
approximation). The radiation pulse can occur in the laboratory frame or in
the atomic rest frame (e.g. when an atomic beam passes through a field
interaction zone). Following the interaction region, the atomic wave
function evolves freely and the kinetic energy term must be included,
leading to such effects as atom focusing and Talbot rephasing, but we
concentrate here on the evolution of the wavefunction only in the field
interaction region.

Even with the simplification afforded by the Raman-Nath approximation, Eqs. (%
\ref{5}) must be solved numerically, in general. To illustrate the relevant
physics, we consider two limiting cases for which an analytic solution can
be obtained, $\sigma _{+},\sigma _{-}$ radiation and $\pi $-polarized
radiation.

\subsection{Basic formalism}

Let us assume that there are a number of ground state manifolds, $G,$ $%
G^{\prime }$, $G^{\prime \prime }$, etc., and that a pair of optical fields
comprising our effective two-photon field drives transitions from sublevels
in one manifold to sublevels in the same or other manifolds. The key point
in considering these Raman transitions is a {\em directionality} in which
one couples the initial state (or states) to the final states by absorption
from field 1 and emission into field 2 but {\em not} by absorption from
field 2 and emission into field 1. Similarly, one couples the final states
to the initial states by absorption from field 2 and emission into field 1
but {\em not} by absorption from field 1 and emission into field 2. Some
simple examples will illustrate how this feat can be accomplished. Suppose
first that there is a single ground state manifold having $G=1$, with the
initial state having $m_{g}=-1$ and the final state having $m_{g}=1$. By
choosing the first field to have $\sigma _{+}$ polarization and the second
field to have $\sigma _{-}$ polarization, the first field couples only the
initial state to the excited states and the second field only the final
state to the excited states. Alternatively, consider initial states in the $%
G $ manifold and final states in the $G^{\prime }$ manifold with $\left|
\omega _{G^{\prime }G}\right| \tau \gg 1$, where $\tau $ is the interaction
time. If one chooses the first field to have frequency $\Omega _{11}$ and
the second field to have frequency $\Omega _{21}$ such that $\left[ \left(
\Omega _{11}-\Omega _{21}\right) -\omega _{G^{\prime }G}\right] \tau
\lesssim 1$ but \label{111}$\left| \Omega _{11}-\Omega _{21}\right| \tau \gg
1,$ then the transition from initial to final states can be realized only by
absorption from field 1 and emission into field 2 (and not by from
absorption from field 2 and emission into field 1).

Consequently, we assume that there are two sets of fields having frequencies 
$\Omega _{1j}$ $\left( j=1,2\ldots n\right) $ and $\Omega _{2j}$ $\left(
j=1,2\ldots n\right) $ from which we can form our required pairs of
two-photon field operators using one field having frequency $\Omega _{1j}$
and the second field having frequency having $\Omega _{2j}.$ To simplify
matters, it is assumed that $\left( \Omega _{1j}-\Omega _{2j}\right) =$%
constant, independent of $j.$ By a proper choice of field frequencies or
polarizations, two-photon field operators formed from any other combination
of fields are assumed to contribute negligibly to the Raman transition
amplitude of interest. Of course there will be two photon operators formed
by using each field with {\em itself }that couples any ground state level to
itself (and possibly to other, degenerate ground state sublevels within the
same manifold). These operators constitute generalized light shift operators
and must be accounted for in the theory, but are not spatially dependent.

In a field interaction representation defined by 
\begin{equation}
a_{Gm_{g}}=e^{i\delta t/2}\tilde{a}_{Gm_{g}},\text{ }a_{G^{\prime
}m_{g}^{\prime }}=e^{-i\delta t/2}\tilde{a}_{G^{\prime }m_{g}^{\prime }}
\label{131}
\end{equation}
where 
\begin{equation}
\delta =\Omega _{1j}-\Omega _{2j}-\omega _{G^{\prime }G},  \label{132}
\end{equation}
one arrives at the evolution equations for the state amplitudes 
\begin{mathletters}
\label{14}
\begin{eqnarray}
i\dot{a}_{Gm_{g}} &=&\left[ \frac{\delta }{2}\delta _{m_{g}m_{g}^{\prime
}}+A_{GG}^{\left( \alpha j,\alpha j\right) }\left( K\right) \left\langle
Gm_{g},KQ|Gm_{g}^{\prime }\right\rangle \varepsilon _{Q}^{K}\left( \alpha
j,\alpha j\right) \right] a_{Gm_{g}^{\prime }}  \nonumber \\
&&+e^{-i{\bf q}_{j}\cdot {\bf r}}\left[ A_{G^{\prime }G}^{\left(
1j,2j\right) }\left( K\right) \left\langle G^{\prime }m_{g}^{\prime
},KQ|Gm_{g}\right\rangle \varepsilon _{Q}^{K}\left( 1j,2j\right) \right]
^{\ast }a_{G^{\prime }m_{g}^{\prime }},  \label{14a} \\
i\dot{a}_{G^{\prime }m_{g}^{\prime }} &=&\left[ -\frac{\delta }{2}\delta
_{m_{g}^{\prime }m_{g}}+A_{G^{\prime }G^{\prime }}^{\left( \alpha j,\alpha
j\right) }\left( K\right) \left\langle G^{\prime }m_{g}^{\prime
},KQ|G^{\prime }m_{g}\right\rangle \varepsilon _{Q}^{K}\left( \alpha
j,\alpha j\right) \right] a_{G^{\prime }m_{g}}  \nonumber \\
&&+e^{i{\bf q}_{j}\cdot {\bf r}}A_{G^{\prime }G}^{\left( 1j,2j\right)
}\left( K\right) \left\langle G^{\prime }m_{g}^{\prime
},KQ|Gm_{g}\right\rangle \varepsilon _{Q}^{K}\left( 1j,2j\right) a_{Gm_{g}},
\label{14b}
\end{eqnarray}
where 
\end{mathletters}
\begin{equation}
{\bf q}_{j}={\bf k}_{1j}-{\bf k}_{2j},  \label{140}
\end{equation}
the tildes have been dropped, and the sum over $\alpha $ is from 1 to 2. The
notation has been changed slightly in that the superscript ($j,j^{\prime }$)
on the $A$'s in Eqs. (\ref{8},\ref{10},\ref{11}) have been replaced by ($%
\alpha j,\alpha ^{\prime }j$).

\begin{center}
\begin{figure}
\begin{minipage}{.5\linewidth}
\begin{center}
\epsfxsize=.95\linewidth \epsfbox{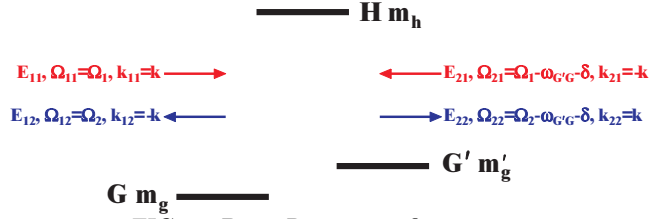}
\end{center}
\end{minipage}
\begin{minipage}{0.99\linewidth} \caption{Basic Raman configuration. 
\label{4pl}}
\end{minipage}
\end{figure}
\end{center}

Consider now the limiting case where there are only two pairs of Raman
fields $j=1,2$ (see Fig. \ref{4pl})$.$%
Looking for a solution to Eqs. (\ref{14}) of the form 
\[
a_{Gm_{g}}=\exp \left[ -i{\bf Q\cdot r/2}\right] \tilde{a}_{Gm_{g}},\text{ \ 
}a_{G^{\prime }m_{g}}=\exp \left[ i{\bf Q\cdot r/}2\right] \tilde{a}%
_{G^{\prime }m_{g}} 
\]
$\left[ {\bf Q=}\left( {\bf q}_{1}+{\bf q}_{2}\right) /2\right] $ and
dropping the tildes, one gets 
\begin{mathletters}
\label{141}
\begin{eqnarray}
i\dot{a}_{Gm_{g}} &=&\left[ \frac{\delta }{2}\delta _{m_{g}m_{g}^{\prime
}}+A_{GG}^{\left( \alpha j,\alpha j\right) }\left( K\right) \left\langle
GK,m_{g}Q|Gm_{g}^{\prime }\right\rangle \varepsilon _{Q}^{K}\left( \alpha
j,\alpha j\right) \right] a_{Gm_{g}^{\prime }}+V_{m_{g}^{\prime
}m_{g}}^{\ast }a_{G^{\prime }m_{g}^{\prime }},  \label{141a} \\
i\dot{a}_{G^{\prime }m_{g}^{\prime }} &=&\left[ -\frac{\delta }{2}\delta
_{m_{g}^{\prime }m_{g}}+A_{G^{\prime }G^{\prime }}^{\left( \alpha j,\alpha
j\right) }\left( K\right) \left\langle G^{\prime }K,m_{g}^{\prime
}Q|G^{\prime }m_{g}\right\rangle \varepsilon _{Q}^{K}\left( \alpha j,\alpha
j\right) \right] a_{G^{\prime }m_{g}}+V_{m_{g}^{\prime }m_{g}}a_{Gm_{g}},
\label{141b}
\end{eqnarray}
where 
\end{mathletters}
\begin{equation}
V_{m_{g}^{\prime }m_{g}}=e^{i{\bf q\cdot r}}A_{G^{\prime }G}^{\left(
11,21\right) }\left( K\right) \left\langle G^{\prime }m_{g}^{\prime
},KQ|Gm_{g}\right\rangle \varepsilon _{Q}^{K}\left( 11,21\right) +e^{-i{\bf %
q\cdot r}}A_{G^{\prime }G}^{\left( 12,22\right) }\left( K\right)
\left\langle G^{\prime }m_{g}^{\prime },KQ|Gm_{g}\right\rangle \varepsilon
_{Q}^{K}\left( 12,22\right)   \label{191}
\end{equation}
and 
\begin{equation}
{\bf q=}\left( {\bf q}_{1}-{\bf q}_{2}\right) /2=\left( {\bf k}_{11}-{\bf k}%
_{21}-{\bf k}_{12}+{\bf k}_{22}\right) /2{\bf .}  \label{192}
\end{equation}

One sees that, under the transformation 
\begin{equation}
{\bf \hat{q}\cdot r}\rightarrow {\bf \hat{q}\cdot r}+\pi /q,  \label{193}
\end{equation}
the off-diagonal elements of the reduced Hamiltonian change their signs. If
the initial density matrix contains no $G-G^{\prime }$ coherence ( for $%
G\neq G^{\prime }$), the final state populations are unaffected by this sign
change. As a consequence atomic state population gratings are produced
having period 
\begin{equation}
d_{g}=2\pi /\left| {\bf k}_{11}-{\bf k}_{21}-{\bf k}_{12}+{\bf k}%
_{22}\right| .  \label{21}
\end{equation}
For ${\bf k}_{1}=-{\bf k}_{1^{\prime }}=-{\bf k}_{2}={\bf k}_{2^{\prime }}=%
{\bf k}$ one gets a $\lambda /4$-period atom grating. In the case of
off-resonant fields, when the $a_{G^{\prime }m_{g}^{\prime }}$ state
amplitude can be adiabatically eliminated, it follows immediately from Eqs. (%
\ref{141}) that the periodicity of the $a_{Gm_{g}}$ state amplitude is $\pi
/q,$ reflecting the fact that atom phase gratings having this periodicity
are produced in the off-resonant case.

\subsection{$\protect\sigma _{+},\protect\sigma _{-}$ radiation}

Let us assume that there is a single ground state manifold having angular
momentum $G=1.$ Atoms interact with a pair of two-photon fields. The first
two-photon field consists of a field \{$E_{11}=E,$ ${\bf k}_{11}={\bf k=}k%
{\bf \hat{z}},$ $\Omega _{11}=\Omega _{1}\}$ which is polarized $\sigma _{+}$
and a field \{$E_{21}=E,$ ${\bf k}_{21}=-{\bf k},$ $\Omega _{21}=\Omega
_{1}-\delta \}$ which is polarized $\sigma _{-}$. This two-photon field is
characterized by a polarization tensor 
\begin{equation}
\varepsilon _{Q}^{K}\left( 11,21\right) =-\delta _{K,2}\delta _{Q,-2}
\label{111}
\end{equation}
and drives transitions between the $m_{g}=\pm 1$ sublevels. The second
two-photon field has the same polarization and field amplitudes, opposite
propagation vectors, ${\bf k}_{12}=-{\bf k}_{22}=-{\bf k,}$ and a carrier
frequency $\Omega _{12}=\Omega _{2}$, which is chosen in a manner to ensure
that $\left| \Omega _{1}-\Omega _{2}\right| \tau \gg 1$, where $\tau $ is
the pulse duration.

For this field polarization, Eqs. (\ref{5}), in the field interaction
representation $a_{1-1}=e^{i\delta t/2}\tilde{a}_{1-1},$ $a_{11}=e^{-i\delta
t/2}\tilde{a}_{11},$ reduce to 
\begin{mathletters}
\label{112}
\begin{eqnarray}
i\dot{a}_{1-1} &=&\left[ \frac{\delta }{2}+S\right] a_{1-1}+T^{\ast }\cos
\left( 2kz\right) a_{11},  \label{112a} \\
i\dot{a}_{11} &=&\left[ -\frac{\delta }{2}+S\right] a_{11}+T\cos \left(
2kz\right) a_{1-1},  \label{112b}
\end{eqnarray}
where the tildes are dropped, 
\end{mathletters}
\begin{equation}
S=A_{11}^{\left( \alpha j,\alpha j\right) }\left( K\right) \left\langle
11,K0|11\right\rangle \varepsilon _{0}^{K}\left( \alpha j,\alpha j\right)
\label{113}
\end{equation}
is a light shift, and 
\begin{equation}
T=-2\left( 3/5\right) ^{1/2}A_{11}^{\left( 11,21\right) }\left( 2\right)
\label{114}
\end{equation}
is a coupling strength$.$

If the atoms are prepared in state $m_{g}=-1$ using optical pumping and if $%
\delta =0$, then a perturbative solution of Eqs. (\ref{112}) yields 
\begin{equation}
\left| a_{11}\right| ^{2}=\left| \int T(t)dt\right| ^{2}\cos ^{2}\left(
2kz\right)  \label{115}
\end{equation}
where the integral is over the pulse duration. The final state population is
modulated with period 
\begin{equation}
d_{g}=\lambda /4.  \label{116}
\end{equation}

\subsection{$\protect\pi -$polarized radiation}

We now return to Raman transitions between $G$ and $G^{\prime }$ manifolds.
For $\pi $-polarized fields $\left( e_{\nu }^{\left( j\right) }=e_{\nu
}^{\left( j^{\prime }\right) }=\delta _{\nu 0}\right) ,$ linearly polarized
along the $z$ axis and propagating in the ${\bf \hat{x}}$ direction, 
\begin{equation}
\varepsilon _{Q}^{K}\left( jj\right) =\varepsilon _{Q}^{K}\left( jj^{\prime
}\right) =3^{-1/2}\left( -\delta _{K0}+2^{1/2}\delta _{K2}\right) \delta
_{Q0},  \label{15}
\end{equation}
and one is led to a collection of independent two-level transitions
characterized by quantum numbers $\left( Gm_{g},G^{\prime }m_{g}\right) $.
Equations (\ref{141}) reduce to 
\begin{mathletters}
\label{16}
\begin{eqnarray}
i\dot{a}_{Gm_{g}} &=&\left( \frac{\delta }{2}+S_{G}\right) a_{Gm_{g}}+\left(
e^{-iqx}T_{1}^{\ast }+e^{iqx}T_{2}^{\ast }\right) a_{G^{\prime }m_{g}},
\label{16a} \\
i\dot{a}_{G^{\prime }m_{g}} &=&\left( -\frac{\delta }{2}+S_{G^{\prime
}}\right) a_{G^{\prime }m_{g}}+\left( e^{iqx}T_{1}+e^{-iqx}T_{2}\right)
a_{Gm_{g}},  \label{16b}
\end{eqnarray}
where $S_{G}$ is the light shift of the initial state given by 
\end{mathletters}
\begin{mathletters}
\label{17}
\begin{equation}
S_{G}=-3^{-1/2}A_{GG}^{\left( \alpha j,\alpha j\right) }\left( 0\right)
+\left( 1-\delta _{G,1/2}\right) \left[ 1.5\left( 2G-1\right) G\left(
G+1\right) \left( 2G+3\right) \right] ^{-1/2}\left[ 3m_{g}^{2}-G\left(
G+1\right) \right] A_{GG}^{\left( \alpha j,\alpha j\right) }\left( 2\right) ,
\eqnum{30}  \label{3.18}
\end{equation}
$S_{G^{\prime }}$ is the light shift of the final state, obtained from (\ref
{3.18}) by the replacement $G\rightarrow G^{\prime },$ and 
\end{mathletters}
\begin{equation}
T_{j}=3^{-1/2}\left[ -\delta _{G^{\prime }G}A_{G^{\prime }G}^{\left(
1j,2j\right) }\left( 0\right) +2^{1/2}\left\langle G^{\prime
}m_{g},20|Gm_{g}\right\rangle A_{G^{\prime }G}^{\left( 1j,2j\right) }\left(
2\right) \right]   \label{18}
\end{equation}
is an effective Rabi frequency for the Raman transition involving absorption
from field $1j$ and emission into field $2j$. For ${\bf k}_{11}=-{\bf k}%
_{11}=-{\bf k}_{21}={\bf k}_{22}={\bf k,}$ and for equal Rabi frequencies
and light shifts, Eqs. (\ref{16}) reduce to (\ref{112}), so that the
solutions discussed below are also relevant for $\sigma _{+}\sigma _{-}$
radiation.

Since the field envelopes are time dependent, the light shifts and Rabi
frequencies are also time dependent, implying that Eq. (\ref{16}) must be
solved numerically, in general. Equations (\ref{16}) can be solved
analytically for rectangular pulses. For pulses having arbitrary shape Eqs. (%
\ref{16}) can be solved analytically in two limiting cases, which we refer
to as ''resonant'' and ''far-detuned''.

In the resonant case, one takes $\delta =0$ and chooses the ratio of the
Rabi frequencies in such a way that $S_{G}=S_{G^{\prime }}=0$ and $%
T_{1}=T_{2}=T.$ Assuming that $T$ is real, one finds for the population of
the final state after the pulse 
\begin{equation}
\left| a_{G^{\prime }m_{g}}\right| ^{2}=\sin ^{2}\left[ \left( \theta
/2\right) \cos \left( qx\right) \right] .  \label{22}
\end{equation}
where $\theta =\int T(t)dt$ is a pulse area.

In the far-detuned case, when the detunings and Rabi frequencies are
sufficiently large on the scale of the inverse pulse duration $\tau ,$%
\begin{equation}
\min \left\{ \delta ,S_{G},S_{G^{\prime }},T_{1},T_{2}\right\} \gg \tau
^{-1},  \label{23}
\end{equation}
it is convenient to use semiclassical dressed states\cite{1}. These states
are obtained by instantaneous diagonalization of Eqs. (\ref{16}). If the
system remains in an instantaneous eigenstate as the field is turned on and
this state adiabatically returns to the initial state following the pulse,
the only modification of the wave function is a phase change of the intial
state probablity amplitudes given by 
\begin{mathletters}
\label{24}
\begin{eqnarray}
a_{Gm_{g}}\left( \infty \right) &=&e^{-i\phi }a_{Gm_{g}}\left( -\infty
\right) ,  \label{24a} \\
\phi &=&\int_{-\infty }^{\infty }dt\left\{ \left[ \frac{1}{4}\left( \delta
+S_{G}-S_{G^{\prime }}\right) ^{2}+\left| T_{1}\right| ^{2}+\left|
T_{2}\right| ^{2}+2\left| T_{1}T_{2}\right| \cos \left( 2qx\right) \right]
^{1/2}-\frac{1}{2}\left( \delta -S_{G}-S_{G^{\prime }}\right) \right\} ,
\label{24b}
\end{eqnarray}
written in the normal interaction represention. In this manner, one creates
a phase grating having period $\lambda /4$ when $q=2k.$

\subsection{$\frac{\protect\lambda }{8}$-period gratings}

\label{lmbd/8}

In the case of single photon transitions, one can create $\frac{\lambda }{4}$%
-period population gratings using counterpropagating traveling wave fields
that are cross polarized. For example, when off-resonant fields drive a $%
G=1/2-H=1/2,$ transition, they produce optical potentials for the $m_{g}=\pm
1/2$ sublevels that have period $\lambda /2$, but are shifted from one
another by $\lambda /4$. If atoms are trapped in these potentials, the
ground state population has period $\lambda /4,$ even though the overall
periodicity of the lattice, including dependence on magnetic state sublevels
remains equal to $\lambda /2.$ In the more general case of arbitrary
detunings of the cross polarized fields, an analysis of Eqs. (\ref{2})
allows one to conclude that excited state population gratings and ground
state population gratings having period $\lambda /4$ can be produced if,
initially, there is no coherence between ground and excited manifolds and
if, in addition, the initial state populations are invariant with respect to
reflection in the $\left( x,y\right) $ plane, i. e. 
\end{mathletters}
\begin{equation}
\left| a_{Gm_{g}}\right| =\left| a_{G,-m_{g}}\right| .  \label{26}
\end{equation}
In the case of two-photon Raman fields having wave vector $2{\bf k}$, one
can anticipate the possibility of creating population gratings having period 
$\lambda /8.$ We now proceed to establish the conditions when this can occur
by considering Eqs. (\ref{141}) with ${\bf q=}2k{\bf \hat{z}}$.

\begin{center}
\begin{figure}
\begin{minipage}{0.9\linewidth}
\begin{center}
\epsfxsize=.95\linewidth \epsfysize=.95\linewidth \epsfbox{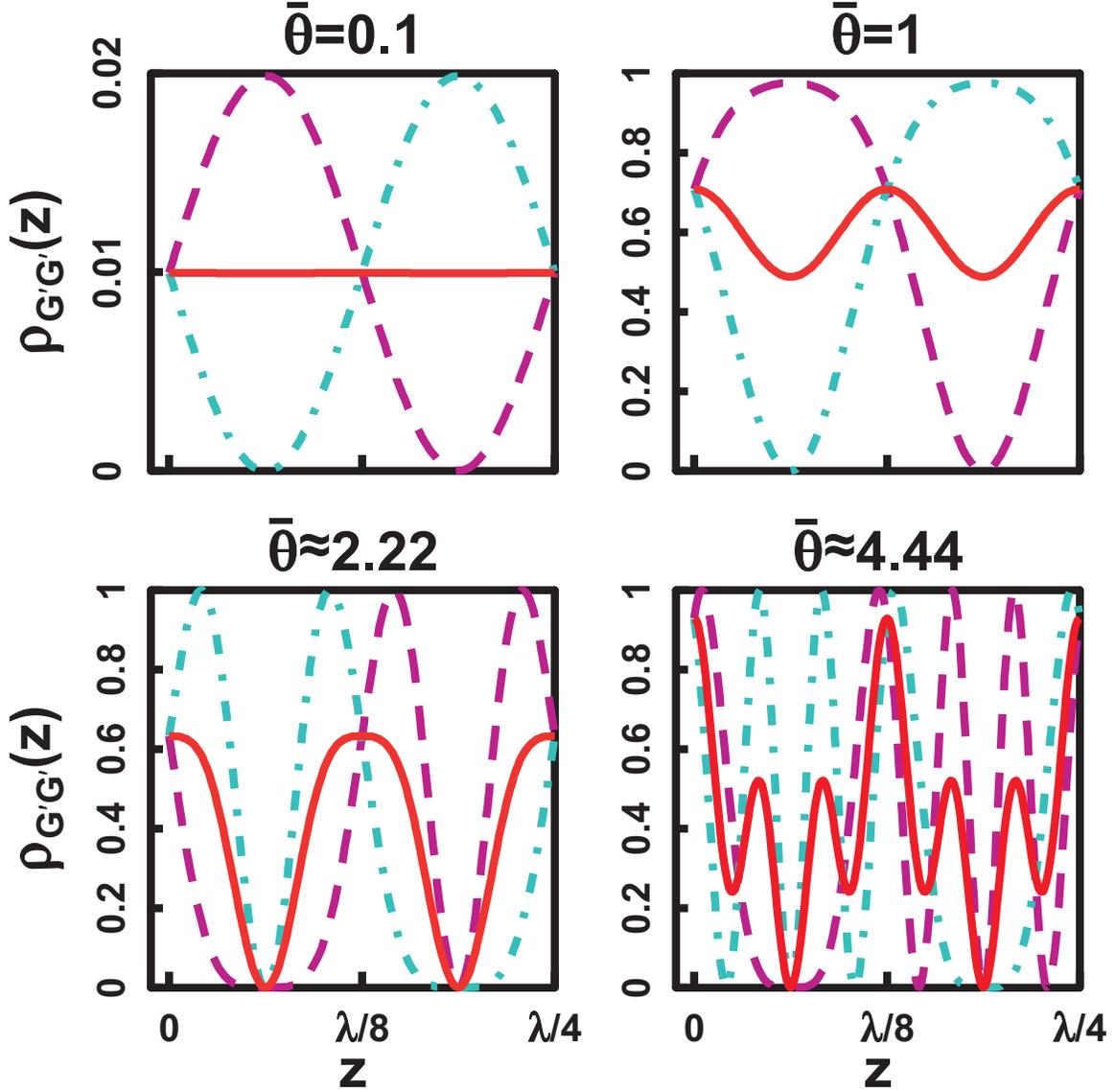}
\end{center}
\end{minipage}
\begin{minipage}{0.99\linewidth} \caption{Population gratings in the $G^{\prime }$ manifold for several
values of $\bar{\protect\theta}.$ The three curves in each graph correspond
to different initial conditions. Solid line: $\protect\rho _{G,\frac{1}{2};G%
\frac{1}{2}}=\protect\rho _{G,-\frac{1}{2};G-\frac{1}{2}}=1/2,$ dashed line: 
$\protect\rho _{G,\frac{1}{2};G\frac{1}{2}}=1,$ dot-dashed line: $\protect%
\rho _{G,-\frac{1}{2};G-\frac{1}{2}}=1,$. 
\label{cx}}
\end{minipage}
\end{figure}
\end{center}

In analogue with single photon transitions, we require that the
transformation $z\rightarrow z+d$ leaves Eqs. (\ref{141}) invariant to
within a global phase factor, along with replacements $a_{G,m_{g}}%
\rightarrow a_{G,-m_{g}}$; $a_{G^{\prime },m_{g}^{\prime }}\rightarrow
a_{G^{\prime },-m_{g}^{\prime }}$. In other words, under the translation $%
z\rightarrow z+d$, the probability amplitude for state $a_{G,-m_{g}}$ as a
function of $z$ is shifted from that of $a_{G,m_{g}}$ by $d,$ to within an
overall phase. These conditions are satisfied, provided Eq. (\ref{26}) is
satisfied and 
\begin{mathletters}
\label{27}
\begin{eqnarray}
\varepsilon _{-Q}^{K}\left( \alpha j,\alpha j\right) &=&\left( -1\right)
^{K}\varepsilon _{Q}^{K}\left( \alpha j,\alpha j\right) ,  \label{27a} \\
\varepsilon _{-Q}^{K}\left( 11,21\right) e^{2ikd} &=&e^{i\phi }\left(
-1\right) ^{K}\varepsilon _{Q}^{K}\left( 11,21\right) \text{; \ }\varepsilon
_{-Q}^{K}\left( 12,22\right) e^{-2ikd}=e^{i\phi }\left( -1\right)
^{K}\varepsilon _{Q}^{K}\left( 12,22\right) .  \label{27b}
\end{eqnarray}
Repeating this transformation, one returns to Eqs. (\ref{141}) for $%
a_{G,m_{g}}$, $a_{G^{\prime },m_{g}^{\prime }}$ at the point $z+2d,$
implying that 
\end{mathletters}
\begin{equation}
e^{2i\phi }=e^{4ikd}=e^{-4ikd}.  \label{28}
\end{equation}
The minimum grating period satisfying this equation is 
\begin{equation}
\lambda _{g}\equiv d_{\min }=\pi /4k=\lambda /8,  \label{30}
\end{equation}
with $\phi =\left( n+\frac{1}{2}\right) \pi ,$ for integer $n$. Equations (%
\ref{27}) are satisfied if 
\begin{mathletters}
\label{31}
\begin{eqnarray}
e_{x}^{\left( \alpha j\right) }e_{y}^{\left( \alpha j\right) } &=&0,
\label{31a} \\
e_{-\nu }^{\left( 11\right) }\left( e_{\nu ^{\prime }}^{\left( 21\right)
}\right) ^{\ast } &=&\left( -1\right) ^{n}e_{\nu }^{\left( 11\right) }\left(
e_{-\nu ^{\prime }}^{\left( 21\right) }\right) ^{\ast },  \label{31b} \\
e_{-\nu }^{\left( 21\right) }\left( e_{\nu ^{\prime }}^{\left( 22\right)
}\right) ^{\ast } &=&\left( -1\right) ^{n+1}e_{\nu }^{\left( 21\right)
}\left( e_{-\nu ^{\prime }}^{\left( 22\right) }\right) ^{\ast },  \label{31c}
\end{eqnarray}
where there is no summation in Eq. (\ref{31a}). From Eq. (\ref{31a}) one
concludes that each field comprising the two-photon Raman field must be
polarized either along ${\bf \hat{x}}$ or along ${\bf \hat{y}}$. Then, one
can verify that Eqs. (\ref{31b}, \ref{31c}) are satisfied for odd $n$ if
fields $\left\{ 11,12\right\} $ are cross-polarized while fields $\left\{
12,22\right\} $ have the same linear polarization,\ along ${\bf \hat{x}}$ or 
${\bf \hat{y}}$.

As an example, we consider the simplest case, $G=G^{\prime }=1/2$ \cite{Sm}.
When ${\bf e}_{11}={\bf e}_{12}={\bf e}_{22}={\bf \hat{x},\,}$and ${\bf e}%
_{21}={\bf \hat{y},}$ one is led to two independent, two-level systems in
which states $\left| G=1/2,1/2\right\rangle \ $and $\left| G^{\prime
}=1/2,1/2\right\rangle $ or states $\left| G=1/2,-1/2\right\rangle $ and $%
\left| G^{\prime }=1/2,-1/2\right\rangle $ are coupled$.$ The state
amplitudes for the first of these evolve according to 
\end{mathletters}
\begin{mathletters}
\label{32}
\begin{eqnarray}
i\dot{a}_{G\frac{1}{2}} &=&\left[ \frac{\delta }{2}-3^{-1/2}A_{GG}^{\left(
\alpha j,\alpha j\right) }\left( 0\right) \right] a_{G\frac{1}{2}}+\left[
-i6^{-1/2}e^{-iqz}\left( A_{G^{\prime }G}^{\left( 11,21\right) }\left(
1\right) \right) ^{\ast }-3^{-1/2}e^{iqz}\left( A_{G^{\prime }G}^{\left(
12,22\right) }\left( 0\right) \right) ^{\ast }\right] a_{G^{\prime }\frac{1}{%
2}},  \label{32a} \\
i\dot{a}_{G^{\prime }\frac{1}{2}} &=&\left[ -\frac{\delta }{2}%
-3^{-1/2}A_{G^{\prime }G^{\prime }}^{\left( \alpha j,\alpha j\right) }\left(
0\right) \right] a_{G^{\prime }\frac{1}{2}}+\left[ i6^{-1/2}e^{iqz}A_{G^{%
\prime }G}^{\left( 11,21\right) }\left( 1\right)
-3^{-1/2}e^{-iqz}A_{G^{\prime }G}^{\left( 12,22\right) }\left( 0\right) %
\right] a_{G\frac{1}{2}}.  \label{32b}
\end{eqnarray}
For the $m_{g},m_{g}^{\prime }=-1/2$ state amplitudes, one has to change the
signs of the terms containing $A_{G^{\prime }G}^{\left( 11,21\right) }\left(
1\right) .$ From this point onwards, we assume that all fields have the same
real pulse envelope function.

For resonant Raman fields ($\delta =0)$, one finds that the populations in
the $G^{\prime }$ manifold following the atom-field interaction are given by 
\end{mathletters}
\begin{mathletters}
\label{33}
\begin{eqnarray}
\rho _{G^{\prime },\pm \frac{1}{2};G^{\prime },\pm \frac{1}{2}} &=&\sin
^{2}\left( \theta _{\pm }\right) \left( \frac{\theta _{1}^{2}+\theta
_{2}^{2}\pm 2\theta _{1}\theta _{2}\sin \left( 4kz\right) }{\theta _{\pm
}^{2}}\right) \rho _{G,\pm \frac{1}{2};G,\pm \frac{1}{2}}^{-};  \label{33a}
\\
\theta _{\pm } &=&\left[ \left( \theta -\theta ^{\prime }\right) ^{2}+\theta
_{1}^{2}+\theta _{2}^{2}\pm 2\theta _{1}\theta _{2}\sin \left( 4kz\right) %
\right] ^{1/2},  \label{33b}
\end{eqnarray}
where $\rho _{G,\pm \frac{1}{2};G,\pm \frac{1}{2}}^{-}$is an initial density
matrix element (it has beeen assumed that initial density matrix elements
involving manifold $G^{\prime }$ vanish), and 
\end{mathletters}
\begin{mathletters}
\label{34}
\begin{eqnarray}
\theta &=&-2^{-1}3^{-1/2}\int dtA_{GG}^{\left( \alpha j,\alpha j\right)
}\left( 0\right) ,  \label{34a} \\
\theta ^{\prime } &=&-2^{-1}3^{-1/2}\int dtA_{G^{\prime }G^{\prime
}}^{\left( \alpha j,\alpha j\right) }\left( 0\right) ,  \label{34b} \\
\theta _{1} &=&6^{-1/2}\int dtA_{G^{\prime }G}^{\left( 11,21\right) }\left(
1\right) ,  \label{34c} \\
\theta _{2} &=&3^{-1/2}\int dtA_{G^{\prime }G}^{\left( 12,22\right) }\left(
1\right)  \label{34d}
\end{eqnarray}
are pulse areas associated with the various two-photon operators. The $%
m_{g}^{\prime }=\pm 1/2$ gratings implicit in Eq. (\ref{33a}) each have
period $\lambda /4,$ and are shifted from one another in space by $\lambda
/8.$ For symmetric initial conditions, $\rho _{G,\pm \frac{1}{2};G,\pm \frac{%
1}{2}}^{-}=1/2,$ one finds that the total population in the $G^{\prime }$
manifold $\rho _{G\prime G\prime }\left( z\right) $ (as well as in the $G$
manifold) is a periodic function of $z$ having period $\lambda /8$.

If the light shifts coincide, i.e. 
\end{mathletters}
\begin{equation}
\theta =\theta ^{\prime },  \label{3401}
\end{equation}
one recovers equations that are identical in from to those for single photon
transitions, except for the reduced periodicity. In this case, for $\rho
_{G,\pm \frac{1}{2};G,\pm \frac{1}{2}}^{-}=1/2$ and 
\begin{equation}
\theta _{1}=\theta _{2}=2^{-1/2}\bar{\theta}  \label{3402}
\end{equation}
one finds the total $G^{\prime }$ manifold population to be 
\begin{equation}
\rho _{G^{\prime }G^{\prime }}\left( z\right) =\sum_{m_{g}^{\prime }=\pm
1/2}\rho _{G\prime m_{g}^{\prime }G\prime m_{g}^{\prime }}\left( z\right)
=0.5\left\{ \sin ^{2}\left[ \bar{\theta}\left( 1-\sin \left( 2qz\right)
\right) ^{1/2}\right] +\sin ^{2}\left[ \bar{\theta}\left( 1+\sin \left(
2qz\right) \right) ^{1/2}\right] \right\} .  \label{341}
\end{equation}
For weak fields, $\bar{\theta}\ll 1,$ the lowest order spatial modulation of
the total population is of order $\bar{\theta}^{4},$ 
\begin{equation}
\rho _{G^{\prime }G^{\prime }}\left( z\right) =\bar{\theta}^{2}-2^{-1}\bar{%
\theta}^{4}\left( 1-3^{-1}\cos \left( 4qx\right) \right) .  \label{342}
\end{equation}
Graphs of $\rho _{G\prime G\prime }\left( z\right) $ for different values of 
$\bar{\theta}$ are shown in Fig. \ref{cx}. Such resonant atom field
interactions are generally not studied in the case of single-photon
transitions, owing to spontaneous decay of the excited state. For Raman
transitions, no such limitations apply.

It is apparent from Fig. \ref{cx} that gratings having contrast approaching
unity can be produced for certain values of pulse area $\bar{\theta}$. One
can approximate values of $\bar{\theta}$ needed to produce near-unity
contrast as follows: From Eq. (\ref{341}), one finds that the population of
each $G^{\prime }$ sublevel {\em vanishes} at $z=\pi /4q$ and . 
\begin{equation}
\bar{\theta}=n\pi /\sqrt{2},  \label{343}
\end{equation}
for integer $n$. On the other hand, the total population at $z=0$ equals
unity when $\bar{\theta}=\left( 2m+1\right) \pi /2$ for integer $m$.
Together with Eq. (\ref{343}), this leads to the requirement 
\begin{equation}
\left( 2m+1\right) /n=\sqrt{2}.  \label{344}
\end{equation}
Though this equation has no solution for integer $m$ and $n,$ one can find a
set of integers, for which Eq. (\ref{344}) is satisfied to arbitrary
accuracy. This set has been generated in Ref. \cite{Chn}. Values of pulse
area (\ref{343}) and grating contrast associated with this set are $\bar{%
\theta}=2.22,$ $4.44,$ $11.1,$ $26.7\ldots ,$ $\rho _{G\prime G\prime
}\left( 0\right) =0.63,$ $0.93,$ $0.988,$ $0.998\ldots $ Graphs of $\rho
_{G\prime G\prime }\left( z\right) $ for two elements of this set are shown
in Fig. \ref{cx}.

In the far detuned case, max$\left\{ \left| \delta \right| ,\text{ }\left|
A_{GG}^{\left( \alpha j,\alpha j\right) }\right| ,\text{ }\left|
A_{G^{\prime }G^{\prime }}^{\left( \alpha j,\alpha j\right) }\right|
\right\} \gg \tau ^{-1},$ instantaneous diagonalization of Eqs. (\ref{32a})
for the $\left| G=1/2,1/2\right\rangle $\ and $\left| G^{\prime
}=1/2,1/2\right\rangle $ states leads to spatially periodic potentials

\begin{eqnarray}
U_{1,2} &=&2^{-1}3^{-1/2}\hbar \left\{ -A_{GG}^{\left( \alpha j,\alpha
j\right) }\left( 0\right) -A_{G^{\prime }G^{\prime }}^{\left( \alpha
j,\alpha j\right) }\left( 0\right) \pm \left[ \left( 3^{1/2}\delta +\left(
A_{G^{\prime }G^{\prime }}^{\left( \alpha j,\alpha j\right) }\left( 0\right)
-A_{GG}^{\left( \alpha j,\alpha j\right) }\left( 0\right) \right) \right)
^{2}\right. \right.  \nonumber \\
&&+\left. \left. 2\left( A_{G^{\prime }G}^{\left( 11,21\right) }\left(
1\right) \right) ^{2}+4\left( A_{G^{\prime }G}^{\left( 21,22\right) }\left(
0\right) \right) ^{2}+2^{5/2}A_{G^{\prime }G}^{\left( 11,21\right) }\left(
1\right) A_{G^{\prime }G}^{\left( 21,22\right) }\left( 0\right) \sin \left(
4kz\right) \right] ^{1/2}\right\} .  \label{35}
\end{eqnarray}
Potentials for the $\left| G=1/2,-1/2\right\rangle \ $and $\left| G^{\prime
}=1/2,-1/2\right\rangle $ states are shifted from these by $\lambda /8.$ For 
$\delta >3^{-1/2}\left( A_{GG}^{\left( \alpha j,\alpha j\right) }\left(
0\right) -A_{G^{\prime }G^{\prime }}^{\left( \alpha j,\alpha j\right)
}\left( 0\right) \right) ,$ the potentials $U_{1}$ are responsible for phase
changes of the initial state amplitudes. In the free evolution following the
atom-field interaction, these atom phase gratings would be converted into
amplitude gratings and the populations at the potential minima would focus
at some specific time following the interaction. Atoms in the $m_{g}=+\frac{1%
}{2}$ Zeeman sublevel focus at $z=3\pi /8k,$ $7\pi /8k$ $\ldots $, while
those in the $m_{g}=-\frac{1}{2}$ focus at $z=\pi /8k,$ $5\pi /8k$ $\ldots $
If both sublevels are equally populated then one obtains a $\lambda /8$%
-period grating of focused atoms that is the analogue of the $\frac{\lambda 
}{4}$-period gratings observed using a single photon transition in Cr atoms 
\cite{Gupta}.

\section{Multicolor fields}

\begin{center}
\begin{figure}
\begin{minipage}{0.5\linewidth}
\begin{center}
\epsfxsize=.95\linewidth \epsfbox{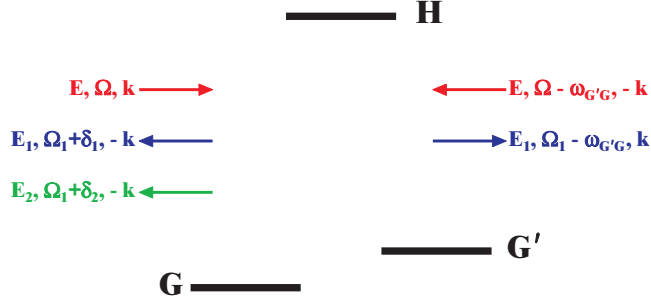}
\end{center}
\end{minipage}
\begin{minipage}{0.99\linewidth} \caption{Multicolor Raman configuration 
\label{5pl}}
\end{minipage}
\end{figure}
\end{center}

In analogy with the multicolor field geometry for single photon transitions 
\cite{ho}, it is possible to suppress low order harmonics in the Raman
scheme by using a geometry involving the three pairs of counterpropagating
fields ${\bf F}=\left\{ {\bf E,E}^{\prime }\right\} ,$ ${\bf F}_{1}=\left\{ 
{\bf E}_{1}{\bf ,E}_{1}^{\prime }\right\} $ and ${\bf F}_{2}=\left\{ {\bf E}%
_{2}{\bf ,E}_{1}^{\prime }\right\} $ shown in Fig. \ref{5pl}, connecting an
initial ground state level $G$ to a final ground state level $G^{\prime }$
via an excited state $H.$ It is assumed that $\left| \Omega _{1}-\Omega
\right| \tau \gg 1.$ Field ${\bf F}$ has effective propagation vector 2{\bf k%
}, while fields ${\bf F}_{1}$ and ${\bf F}_{2}$ have effective propagation
vector $-2{\bf k}${\bf .}

\begin{center}
\begin{figure}
\begin{minipage}{0.5\linewidth}
\begin{center}
\epsfxsize=.95\linewidth \epsfbox{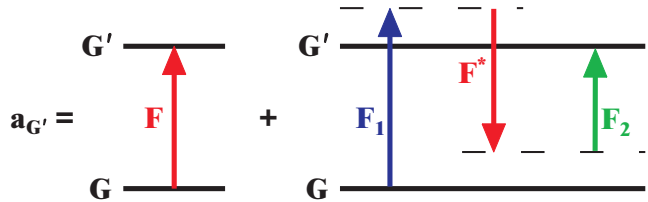}
\end{center}
\end{minipage}
\begin{minipage}{0.99\linewidth} \caption{Contributions to $G\rightarrow G^{\prime }$ transition amplitude
when $\protect\delta _{1}=-\protect\delta _{2}.$ 
\label{6pl}}
\end{minipage}
\end{figure}
\end{center}

When the Raman detunings are large, $\left| \delta _{j}\right| \tau \gg 1,$
basic Raman processes are suppressed; however, by choosing $n_{1}\delta
_{1}+n_{2}\delta _{2}=0$, where $n_{1}$ and $n_{2}$ are positive integers,
one can produce nearly sinusoidal, high-order gratings. For example, if $%
\delta _{1}=-\delta _{2}$ and $\delta =0,$ to lowest order in the atom field
coupling, there are two {\em resonant} contributions to the $G\rightarrow
G^{\prime }$ transition amplitude, shown schematically in Fig. \ref{6pl}.
One contribution is associated with a two-quantum process and varies as $%
\exp \left( 2ikz\right) $ while the other is associated with a six-quantum
process and varies as $\exp \left( -6ikz\right) .$ In contributing to the
final state probability, these terms interfere and result in a $\frac{%
\lambda }{8}$-period population grating. For arbitrary $n_{1}$ and $n_{2}$
one produces a $\frac{\lambda }{4\left( n_{1}+n_{2}\right) }$-period
grating. By choosing $\left| \delta \right| \tau \gg 1$, one can produce
high-order phase gratings. The higher order gratings can simultaneously have
near unity contrast and nearly sinusoidal shape\cite{ho}. This is an
important advantage of the multicolor field technique over the basic Raman
configuration if one wishes to produce nearly sinusoidal, high-order, high
contrast atom gratings.

The resulting equations are completely analogous to those obtained
previously for the case of single-photon transitions \cite{ho} and will not
be repeated here. There are two important differences between the two cases,
aside from the fact that single-photon Rabi frequencies are replaced by
two-photon Rabi frequencies. In order to satisfy the adiabaticity
requirements necessary to suppress the lower harmonics, it is essential that
the detunings $\left| \delta _{j}\right| $ be larger than any relevant
characteristic frequencies in the problem such as decay rates or relative
light shifts. In the case of single-photon transitions this led one to
choose the field intensities such that $\chi _{\alpha }^{2}/\Delta _{1}+\chi
_{\beta }^{2}/\Delta _{2}=0$, where $\chi _{\alpha },\chi _{\beta },\Delta
_{1},\Delta _{2}$ are single photon Rabi frequencies and detunings, and to
limit the pulse durations to values for which $\gamma _{e}\tau \ll 1,$ where 
$\gamma _{e}$ is an excited state decay rate \cite{ho}. The limitation
placed on the pulse duration necessitates the use of stronger fields to
reach higher order harmonics \cite{ho}. In the case of ground state
transitions, there is no longer any significant restriction on pulse
duration since the initial and final states are long-lived. However, the
requirements on the relative light shifts remains the same with the
replacement of single photon Rabi frequencies and detunings by the
corresponding two-photon Rabi frequencies and detunings; moreover, the
single photon detunings $\Delta _{H,G}^{\left( j\right) }$ should be chosen
to cancel the {\em first-order} relative light shifts.

It is also possible to use an alternative approach to produce higher order
gratings. Cataliotti {\it et al.} \cite{cat} detuned their Raman fields such
that $\delta =\omega _{G^{\prime }G}/n$, where $n$ is a positive integer,
and observed transitions for $n$ as large as 25. In other words, they
monitored a multiphoton Raman transition, requiring $n$ Raman fields to
achieve resonance. They used copropagating fields, but if the copropagating
fields are replaced by counterpropagating fields, the multiphoton Raman
field acts as a traveling wave field having propagation vector $2n{\bf k}$.
If a second Raman field having a different carrier frequency and an
effective propagation vector -$2n{\bf k}$ is added, one creates a ''standing
wave'' multiphoton Raman field varying as $\cos (2n{\bf k\cdot r)}$. In this
manner one can create atom gratings having periodicity $\lambda /4n$.

\section{Discussion}

It has been shown that it is possible to use optical fields having
wavelength $\lambda $ to create atom amplitude and phase gratings having
period $\lambda /4$ and $\lambda /8$ using a basic Raman geometry, or $%
\lambda /4(n_{1}+n_{2})$ using a multicolor field geometry. Once the
gratings are created, the question remains as how to image the gratings at
some distance $L$ from the atom-field interaction zone. This question has
been addressed in detail in a previous publication \cite{ho} for both
amplitude and phase gratings. For highly collimated beams, the phase
gratings evolve into a focused array of lines having spacing $\lambda
/4(n_{1}+n_{2})$ which could be deposited on a substrate. For atom beams
having a higher angular divergence, echo techniques can be used to generate
gratings with even smaller periodicities at specific focal planes \cite{ho}.
As such, the basic and multicolor Raman geometries offer interesting
possibilities for atom nanofabrication.

It is interesting to return to Eqs. (\ref{14}) for {\em cw} fields. As
mentioned earlier, the eigenvalues of the Hamiltonian correspond to the
optical potentials for the ground state manifold. Since these optical
potentials have period $\lambda /4$ or smaller, the normal or multicolor
Raman field geometry can be used to produce optical lattices having this
reduced periodicity. As an illustrative example, we consider the level
scheme of Sec. \ref{lmbd/8} involving two ground state manifolds with $G=1/2$
and $G^{\prime }=1/2$ that are connected to an excited state by four fields
in the basic Raman geometry; \{$E_{11},$ $\Omega _{11}=\Omega _{1},$ ${\bf k}%
_{11}={\bf k,}$ ${\bf e}_{11}={\bf \hat{x}\}}$, \{$E_{21},$ $\Omega
_{21}=\Omega _{1}-\omega _{G^{\prime }G}-\delta ,$ ${\bf k}_{21}=-{\bf k,}$ $%
{\bf e}_{21}={\bf \hat{y}\}}$, \{$E_{12},$ $\Omega _{12}=\Omega _{2},$ ${\bf %
k}_{12}=-{\bf k,e}_{12}={\bf \hat{x}\}}$, \{$E_{2},$ $\Omega _{22}=\Omega
_{2}-\omega _{G^{\prime }G}-\delta ,$ ${\bf k}_{22}={\bf k,}$ ${\bf e}_{22}=%
{\bf \hat{x}\}}.$ The corresponding optical potentials produced are given by
Eq. (\ref{35}). If atoms are prepared in the $G$ manifold and $\delta $ is
sufficiently large to enable one to adiabatically eliminate the $G^{\prime }$
state amplitudes, one finds that the ground state amplitudes in the $G$
manifold evolve as 
\begin{eqnarray*}
a_{G,\pm \frac{1}{2}} &\propto &\exp \left( -iU_{\pm }t/\hbar \right)
a_{G,\pm \frac{1}{2}}^{-}, \\
U_{\pm } &=&\pm 2^{1/2}\hbar \left( A_{G^{\prime }G}^{\left( 11,21\right)
}\left( 1\right) A_{G^{\prime }G}^{\left( 12,22\right) }\left( 0\right)
/3\delta \right) \sin \left( 4kz\right)
\end{eqnarray*}
In other words, the $m_{g}=1/2$ sublevel is subjected to the $U_{+}$
potential and the $m_{g}=-1/2$ sublevel is subjected to the $U_{-}$
potential. If cold atoms are trapped in these potentials the atomic density
will have $\lambda /8$ periodicity. Even smaller periodicities are possible
using a multicolor geometry. Calculations of the optical potentials for more
complicated level schemes and field geometries, such as those appropriate to
the alkali metal atoms, are deferred to a future planned publication.

\section{Acknowledgments}

The extension of our previous work on multicolor geometries to the Raman
case was suggested to us by Tycho Sleator at New York University. We are
pleased to acknowledge helpful discussions with G. Raithel at the University
of Michigan. This work is supported by the U. S. Office of Army Research
under Grant No. DAAD19-00-1-0412 and the National Science Foundation under
Grant No. PHY-9800981, Grant No. PHY-0098016, and the FOCUS Center Grant,
and by the Office of the Vice President for Research and the College of
Literature Science and the Arts of the University of Michigan.


\begin{references}
\bibitem{singlecoupling}  For the analogue with single photon transitions to
be correct, it is necessary that the Raman fields obey a directionality
property that is discussed at the beginning of Sec. III A.
\bibitem{Sm}  There are only a few atoms which have hyperfine sublevels $%
G=G^{\prime }=1/2$ belonging to different fine structure manifolds. One such
atom is Sm, for which the Raman transition $\left|
^{7}F_{3},G=1/2\right\rangle \rightarrow \left| ^{7}D_{3},H\right\rangle
\rightarrow \left| ^{7}F_{4},G^{\prime }=1/2\right\rangle ,$ can be driven
by optical radiation at $\lambda =473$ nm and $\lambda =491$ nm.
\end{references}
\end{document}